\documentclass[10pt,a4letter]{article}
\usepackage[english]{babel}
\usepackage{comment}

\usepackage{amsmath}
\usepackage{amssymb}
\usepackage{graphicx,epsfig}
\usepackage[mathscr]{eucal}

\usepackage{color}
\usepackage[numbers,sort&compress]{natbib}
\usepackage{multirow}
\usepackage{tabularray}
\usepackage{tabularx}
\UseTblrLibrary{booktabs}
\usepackage{float}

\usepackage{slashed}
\usepackage{braket}

\usepackage{scalefnt,ulem,pstricks}

\usepackage{xspace}
\usepackage{setspace}
\usepackage{xstring}
\usepackage{mathrsfs}
\usepackage{amsbsy}
\usepackage{makecell}

\usepackage{lipsum}
\usepackage[colorlinks=false,filecolor=black,urlcolor=cyan]{hyperref}
\usepackage[capitalise]{cleveref}

\newcommand\as{\alpha_{\mathrm{S}}}
\newcommand\muR{\mu_{\rm R}}
\newcommand\muF{\mu_{\rm F}}
\newcommand\muIR{\mu_{\rm IR}}
\newcommand\M{{\cal M}}

\newcommand\Matrix{{\sc Matrix}\xspace}
\newcommand\OpenLoops{{\sc OpenLoops}\xspace}

\newcommand{\eps}{\epsilon}

\def\ttH{\ensuremath{t \bar t H}\xspace}
\def\ttW{\ensuremath{t \bar t W}\xspace}
\def\ttWm{\ensuremath{t \bar t W^-}\xspace}
\def\ttWp{\ensuremath{t \bar t W^+}\xspace}
\def\tttt{\ensuremath{t {\bar t} t {\bar t}}\xspace}

\begin{document}
\begin{titlepage}
\begin{flushright}
ZU-TH 20/26 \\
TUM-HEP-1604/26 \\
MPP-2026-108\\
CERN-TH-2026-131
\end{flushright}

\vspace*{0.5cm}

\begin{center}
  {\Large \bf NNLO QCD predictions for $t\bar t W$ production at hadron colliders}
\end{center}

\par \vspace{2mm}
\begin{center}

 {\bf Matteo~Becchetti${}^{(a)}$}, {\bf Dhimiter~Canko${}^{(a)}$}, {\bf Xiang~Chen${}^{(b)}$}, {\bf Vsevolod~Chestnov${}^{(c)}$}, \\[0.2cm]
 {\bf Maximilian~Delto${}^{(d)}$}, {\bf Sara~Ditsch${}^{(e,f)}$}, {\bf Massimiliano~Grazzini${}^{(b)}$}, {\bf Stefan~Kallweit${}^{(b)}$}, \\[0.2cm]
 {\bf Tiziano~Peraro${}^{(a)}$}, {\bf Mattia~Pozzoli${}^{(a)}$}, {\bf Chiara~Savoini${}^{(f)}$}, {\bf Lorenzo~Tancredi${}^{(f)}$}\\[0.2cm]
 and {\bf Simone~Zoia${}^{(b)}$}

\vspace{5mm}
${}^{(a)}$ Dipartimento di Fisica e Astronomia, Universit\`{a} di Bologna and\\ INFN, Sezione di Bologna, via Irnerio 46, 40126 Bologna, Italy\\[0.25cm]
${}^{(b)}$ Physik-Institut, Universit\"at Z\"urich, Winterthurerstrasse 190, 8057 Z\"urich, Switzerland\\[0.25cm]
${}^{(c)}$ Mathematical Institute, University of Oxford, OX2 6GG, United Kingdom\\[0.25cm]
${}^{(d)}$ Theoretical Physics Department, CERN, 1211 Geneva 23, Switzerland\\[0.25cm]
${}^{(e)}$ Max-Planck-Institut f\"ur Physik, Werner-Heisenberg-Institut, Boltzmannstraße 8, 85748 Garching, Germany \\[0.25cm]
${}^{(f)}$ Technical University of Munich, TUM School of Natural Sciences, Physics Department, James-Franck-Straße 1, 85748 Garching, Germany\\[0.25cm]

\end{center}

\par \vspace{2mm}
\begin{center} {\large \bf Abstract}

\end{center}
\begin{quote}
\pretolerance 10000

The production of a top--antitop quark pair in association with a $W$ boson constitutes one of the heaviest final states currently studied at the Large Hadron Collider (LHC) at CERN. Measurements of its production rate have consistently exceeded Standard Model predictions. Owing to the complexity of the two-loop amplitudes entering the double-virtual correction, next-to-next-to-leading-order (NNLO) QCD calculations for this process have so far employed dynamical approximations for the two-loop contribution. We present NNLO QCD predictions based, for the first time, on a direct computation of the required two-loop amplitudes in the generalised leading-colour limit.

\end{quote}

\vspace*{\fill}
June 2026
\end{titlepage}

\section{Introduction}

The associated production of a top–antitop quark pair and a $W$ boson ($\ttW$) provides one of the most intriguing massive signatures currently accessible at the LHC. At leading order (LO) in the QCD coupling, this process is induced by the scattering of a quark--antiquark pair with different flavours, while gluon channels open up only at higher orders. This leads to a charge-asymmetric production between $\ttWp$ and $\ttWm$, and to an intricate pattern of QCD and electroweak (EW) radiative corrections.
Beyond its intrinsic interest as a rare process, a precise understanding of $\ttW$ production is crucial because it constitutes an important background to many searches and measurements at the LHC. In particular, as one of the few Standard Model (SM) processes yielding same-sign lepton pairs, it represents the leading irreducible background to analyses exploiting this signature, such as the production of four top quarks (\tttt) or the associated production of a top--antitop pair with a Higgs boson (\ttH). The rarity of these final states also makes them a powerful probe of physics beyond the SM~\cite{ATLAS:2018alq,ATLAS:2019fag,CMS:2020cpy}.

The inclusive \ttW cross section has been directly measured by the ATLAS and CMS collaborations
at centre-of-mass energies of \mbox{$\sqrt{s}=8$\,TeV}~\cite{ATLAS:2015qtq,CMS:2015uvn,CMS:2022tkv} and
\mbox{$\sqrt{s}=13$\,TeV}~\cite{ATLAS:2016wgc,CMS:2017ugv,ATLAS:2019fwo,ATLAS:2023gon,ATLAS:2024moy,CMS:2025iwa}.
The measured rates have generally been found to exceed the corresponding SM predictions, and
a similar pattern is observed in indirect \ttW measurements within \ttH~\cite{ATLAS:2019nvo,CMS:2020mpn} and \tttt~\cite{CMS:2023ftu,ATLAS:2023ajo} analyses.

In view of the persistent excess of the measured \ttW rates over the corresponding SM predictions, increasingly precise theoretical calculations are required. This has motivated a continuous effort to improve the perturbative description of \ttW production, including higher-order QCD and EW corrections.
Next-to-leading order~(NLO) QCD corrections to on-shell $\ttW$ production have been
presented in Refs.~\cite{Badger:2010mg,Campbell:2012dh,Maltoni:2015ena}, while NLO EW contributions
were evaluated in Refs.~\cite{Frixione:2015zaa,Frederix:2017wme}.
Soft-gluon effects were investigated in Refs.~\cite{Li:2014ula,Broggio:2016zgg,Kulesza:2018tqz,Broggio:2019ewu}.
NLO corrections to the complete off-shell $\ttW$ process were considered first in QCD \cite{Bevilacqua:2020pzy,Denner:2020hgg,Bevilacqua:2020srb}, and then in QCD+EW \cite{Denner:2021hqi}.
NLO QCD+EW predictions were also supplemented with multi-jet merging~\cite{Frederix:2012ps,Frederix:2021agh}.

In 2023, the first NNLO QCD predictions for the inclusive $\ttW$ cross section were presented~\cite{Buonocore:2023ljm}.
That calculation relied on two different approximations of the two-loop amplitude, where either the $W$ boson is considered soft or the top-quark mass small compared to all other scales.
Although a conservative estimate of the uncertainties associated with these approximations indicated that their impact is subleading compared to the residual perturbative uncertainties, an exact computation of the two-loop contribution is ultimately required to confirm the results of Ref.~\cite{Buonocore:2023ljm} and to eliminate the systematic uncertainties due to the adopted approximations.

In this Letter we take a crucial step in this direction by presenting an NNLO QCD calculation for $\ttW$ production in which the two-loop amplitude is evaluated, for the first time, in the generalised leading-colour (LC) limit.
This approximation is widely used in the community and, unlike those adopted in Ref.~\cite{Buonocore:2023ljm}, is controlled by a fixed expansion parameter.
The complex \mbox{$2\to 3$} kinematics, paired with multiple masses in both external and internal legs, gives rise to an extreme algebraic complexity and to analytic structures for which a satisfactory mathematical formalism is still missing, placing this calculation among the most demanding applications of modern amplitude techniques.
We provide new NNLO predictions based on the new two-loop amplitude results, estimate their uncertainties and compare our results with those of Ref.~\cite{Buonocore:2023ljm}.

\section{Calculation}

The cross section is computed within the \Matrix framework \cite{Grazzini:2017mhc}, suitably extended to $\ttW$ production \cite{Buonocore:2023ljm}, following the work done for $t{\bar t}$ \cite{Catani:2019iny,Catani:2019hip}, $b{\bar b}$ \cite{Catani:2020kkl}, $b{\bar b}W$ \cite{Buonocore:2022pqq} and $t{\bar t}H$~\cite{Devoto:2024nhl} production.
NNLO infrared~(IR) singularities are handled and cancelled by using the $q_T$-subtraction formalism~\cite{Catani:2007vq}, extended to heavy-quark production in Refs.~\cite{Bonciani:2015sha,Catani:2019iny,Catani:2019hip}. NLO IR singularities in the \mbox{$\ttW+{\rm jet}$} contribution entering the NNLO corrections are cured by applying the dipole subtraction formalism~\cite{Catani:1996jh,Catani:1996vz,Catani:2002hc}.
In the case of heavy-quark production and related processes, additional soft-parton contributions have to be included \cite{Catani:2023tby}, which are now available for general kinematics in the {\sc Shark} package \cite{Devoto:2025eyc}.
All tree-level and one-loop amplitudes required in the calculation are obtained
with \OpenLoops~\cite{Cascioli:2011va, Buccioni:2017yxi,Buccioni:2019sur}.
The finite part of the $n$-loop virtual amplitude contributes to the N$^n$LO correction in the form
\begin{equation}
\label{eq:sigmaH}
\mathrm{d} \sigma_{H^{(n)}}=\left(\frac{\as(\muR)}{4\pi}\right)^n H^{(n)}\times \mathrm{d}\sigma_{\rm LO} \,,
\end{equation}
where $\as(\muR)$ is the QCD coupling at the renormalisation scale $\muR$, $\mathrm{d}\sigma_{\rm LO}$ is the Born-level cross section, and the hard-virtual coefficient $H^{(n)}$ is defined as
\begin{equation}
\label{eq:Hn}
H^{(n)}(\muIR) = \frac{2 \, {\rm Re}\left[\M^{(n)}_{\rm fin}(\muR, \muIR) \, \M^{(0)*} \right]}{\left|\M^{(0)}\right|^2}\biggl|_{\muR=Q}\,.
\end{equation}
In Eq.~(\ref{eq:Hn}), $\M$ is the renormalised scattering amplitude for the process ${\bar u}d (\bar{d} u)\to t{\bar t}W^{-(+)}$, and the sum over polarisations is understood.
The superscript `$(n)$' denotes the $n$-th perturbative order in an expansion in powers of $\as(\mu_R)/(4\pi)$,
while the subscript `${\rm fin}$' indicates the finite part, upon subtraction of IR singularities according to the prescription of Ref.~\cite{Ferroglia:2009ii} at the scale $\muIR=Q$, with $Q$ the invariant mass of the $\ttW$ system.

The calculation of the two-loop amplitude $\M^{(2)}$ is the main novelty of this work.
We will give a detailed discussion in a separate article and outline the main aspects here.
We follow the conventions of Ref.~\cite{Becchetti:2025osw}, where the one-loop amplitude for the same process has been computed up to $\mathcal{O}(\eps^2)$ in the dimensional regulator $\eps$.
We consider the partonic process
\begin{align} \label{eq:processes}
\bar{u}(p_1) d(p_2) \bar{t}(p_3) t(p_4) W^+(p_5) \rightarrow 0 \, .
\end{align}
Since the $W$ boson only couples to the light-quark line and the vertex is insensitive to the quark type, the amplitude for the process in \cref{eq:processes} is identical to that for $\bar{d}(p_1) u(p_2) \bar{t}(p_3) t(p_4) W^-(p_5) \rightarrow 0$.
The kinematics is described by seven Lorentz invariants and one pseudo-scalar invariant.
We generate the Feynman diagrams with \texttt{QGRAF}~\cite{Nogueira:1991ex}, and use \texttt{FORM}~\cite{Ruijl:2017dtg,Davies:2026cci} to insert the Feynman rules and perform the colour algebra.
We work in the LC limit, i.e.\ we keep only the terms of the two-loop amplitude proportional to $N_c^2$, $N_c n_l$ and $n_l^2$, where \mbox{$N_c=3$} is the number of colours and \mbox{$n_l=5$} the number of light quarks circulating in the loops.
Following Ref.~\cite{Becchetti:2025osw}, we decompose the bare amplitude $\M^{(2)}_{\rm bare}$ into 24 independent tensor structures $\mathcal{T}_i$ by means of physical projectors~\cite{Peraro:2019cjj,Peraro:2020sfm} in the `t Hooft-Veltman scheme~\cite{tHooft:1972tcz}, and compute the amplitude projections,
\begin{equation} \label{eq:preFF}
\mathcal{B}^{(2)}_i = \sum_{\rm{pol.}} \mathcal{T}_i^\dagger \M^{(2)}_{\rm bare} \,.
\end{equation}
At this stage, the projections are linear combinations with rational coefficients of scalar Feynman integrals of the families studied in Ref.~\cite{Becchetti:2025qlu} and crossings thereof.
The goal is to express them as
\begin{align} \label{eq:preFFexpr}
\mathcal{B}^{(2)}_i(X, \eps) = \sum_{k=-4}^0  \sum_{j} \eps^k \, c_{ikj}(X) \, F_j(X) + \mathcal{O}(\eps) \,,
\end{align}
where $c_{ikj}$ are linearly independent rational functions of the kinematic invariants $X$, and $F_j$ are polynomials of special functions and transcendental constants ($\zeta_2$ and $\zeta_3$), as well as square roots coming from the definition of the integral bases.
In order to reach the form in \cref{eq:preFFexpr}, we faced two main obstacles.

First, the relevant Feynman integrals involve complicated analytic structures, including elliptic curves and nested square roots~\cite{Becchetti:2025qlu}, which make it challenging to obtain full numerical control over the phase space of a $2\to3$ process.
In order to circumvent this issue, we follow the approach of Refs.~\cite{Badger:2024dxo,Badger:2025ljy}. We expand the integral bases of Ref.~\cite{Becchetti:2025qlu} around \mbox{$\eps=0$}, and express the
coefficients as polynomials in a (potentially) over-complete set of special functions, defined as the solution to a system of algebraic partial differential equations (DEs).
We evaluate the special functions by solving these DEs numerically with \textsc{AMFlow}'s DE solver~\cite{Liu:2022chg}, starting from $30$-digit boundary values in the physical region, also obtained with \textsc{AMFlow}~\cite{Liu:2017jxz,Liu:2022chg}.
Despite the leap in complexity compared to the case studied in Refs.~\cite{Badger:2024dxo,Badger:2025ljy}, this approach
still yields a substantial simplification of the expression of the amplitude, and allows us to analytically subtract the ultraviolet (UV) and IR poles to obtain the finite remainder.
To this end, all ingredients entering the subtraction must be written in terms of the same set of special functions.
Therefore, we employed our setup to recompute the LC one-loop amplitude as well as the one-loop amplitude with the insertion of a mass counterterm.
We validated the former against Ref.~\cite{Becchetti:2025osw}.
In this way, the projections of the two-loop finite remainder take the form
\begin{align} \label{eq:preFFfin}
\mathcal{B}^{(2)}_{{\rm fin},i}(X) = \sum_{j} r_{ij}(X) \, F_j(X) \,,
\end{align}
where the rational coefficients $r_{ij}(X)$ come from either the tree-, one- or two-loop amplitude.
The hard-virtual coefficient $H^{(2)}$ is given by a linear combination of the projections $\mathcal{B}^{(2)}_{{\rm fin},i}$.

The second obstacle is the algebraic complexity of the expressions throughout the calculation.
To overcome this, we perform all rational operations on rational functions numerically over finite fields~\cite{vonManteuffel:2014ixa,Peraro:2016wsq} within a \textsc{FiniteFlow} dataflow graph~\cite{Peraro:2019svx}.
The main steps are the solution of the integration-by-parts (IBP) relations~\cite{Tkachov:1981wb,Chetyrkin:1981qh,Laporta:2000dsw} generated with \textsc{NeatIBP}~\cite{Wu:2023upw} to reduce the scalar integrals to the integral bases of Ref.~\cite{Becchetti:2025qlu}, and the Laurent expansion around \mbox{$\eps = 0$}.
Due to the complexity of the coefficients $c_{ikj}(X)$, instead of reconstructing their analytic expressions, we lift their numerical values from finite fields to rational numbers point by point, following the approach illustrated in Ref.~\cite{Peraro:2019okx}.
For each target phase-space point, this procedure requires combining finite-field evaluations over multiple primes through the Chinese remainder theorem.
The number of needed primes depends not only on the complexity of the coefficients, but also on the values of the invariants, which need to be rationalised.
More details on this will be given in a dedicated publication.
Here, we limit ourselves to reporting that the maximum number of primes needed for this calculation is 400.
At the same phase-space point, reconstructing the values of the $\eps$-expanded rational coefficients of the basis integrals would have instead required 608 primes.
This, combined with the analytic cancellation of poles, motivates our representation of the basis integrals in terms of special functions even in the context of a numerical calculation.
Despite the large number of required prime fields, this approach is effective in this context because it removes any numerical error from the evaluation of the coefficients without severely compromising performance, as the evaluation of the special functions remains the main bottleneck.

We validate our results for the two-loop finite remainder with an independent numerical calculation.
In this approach, we compute directly the unpolarised interference with the tree-level amplitude, and
adopt the conventional dimensional regularisation scheme, as opposed to the 't Hooft-Veltman scheme.
For the integral reduction, we employ \textsc{Blade}~\cite{Guan:2024byi} to construct a block-triangular linear system of IBP relations~\cite{Guan:2019bcx}.
A novel feature of our approach is that this linear system is solved numerically in \textsc{Mathematica} at each phase-space point using floating-point values for the basis integrals.
The latter are evaluated by solving the DEs of Ref.~\cite{Becchetti:2025qlu} with \textsc{AMFlow}'s DE solver.
In addition, we set not only the kinematic variables and masses, but also $\eps$ to numerical values, specifically $\eps = \pm \bar \epsilon$ where $\bar \epsilon = 1/1000$, throughout the entire calculation~\cite{Bi:2023bnq,Bi:2025oga,Li:2025bsq}.
By subtracting UV/IR poles and combining the finite parts computed with the two values of $\eps$, we determine the finite remainder with an uncertainty of order $\mathcal{O}(\bar\epsilon^2)$.
Finally, we cross-check the two-loop finite remainder against the value derived as discussed above at three representative phase-space points and find agreement within five significant digits, as expected, thereby providing a robust check of the entire computation.

We evaluate the two-loop finite remainder on a five-dimensional grid of 224'640 phase-space points generated according to the parametrisation of Ref.~\cite{Agarwal:2024jyq} in terms of two energy fractions and three angles.
We observe large cancellations and loss of precision at a small subset of points.
For those points we have re-evaluated the special functions to higher precision in order to ensure at least five correct digits at the level of $\M^{(2)}_{\rm fin}(\muR, \muIR) \, \M^{(0)*}$ throughout the grid.
The two-loop hard-virtual coefficient $H^{(2)}$ is eventually evaluated using a quadratic B-spline interpolation through the {\sc Bsplines-fortran} package~\cite{bsplines}.
In principle, similar results with fewer grid points may be obtained with Machine Learning methods (see e.g.\ Ref.~\cite{Breso-Pla:2024pda}), but we do not explore this possibility here.

\section{Results}

We now present our numerical results for a centre-of-mass energy of \mbox{$\sqrt{s}=13$\,TeV}.
To directly compare with the calculation of Ref.~\cite{Buonocore:2023ljm}, we adopt the same setup.
The pole mass of the top quark is set to \mbox{$m_t = 173.2$\,GeV}, and the $W$-boson mass to \mbox{$m_W = 80.385$\,GeV}.
We work in the $G_\mu$-scheme for the EW parameters,
with \mbox{$G_\mu = 1.16639\times 10^{-5}$\,GeV$^{-2}$} and \mbox{$m_Z = 91.1876$\,GeV}.
The CKM matrix is taken to be diagonal.
We use the \verb|NNPDF31_nnlo_as_0118_luxqed| set for parton
distribution functions~\cite{Bertone:2017bme} and strong coupling, through the \verb|LHAPDF| interface~\cite{Buckley:2014ana}.
For our central predictions, we set the
renormalisation~($\muR$) and factorisation~($\muF$) scales to 
\mbox{$m_t + m_W/2$}, and evaluate the perturbative uncertainties through a standard seven-point scale variation, i.e.\ we independently vary $\muF$ and $\muR$ by a factor of two with the constraint \mbox{$1/2 \leq \muR/\muF \leq 2$}.

\begin{figure}[t]
  \centering
  \includegraphics[width=0.6\textwidth]{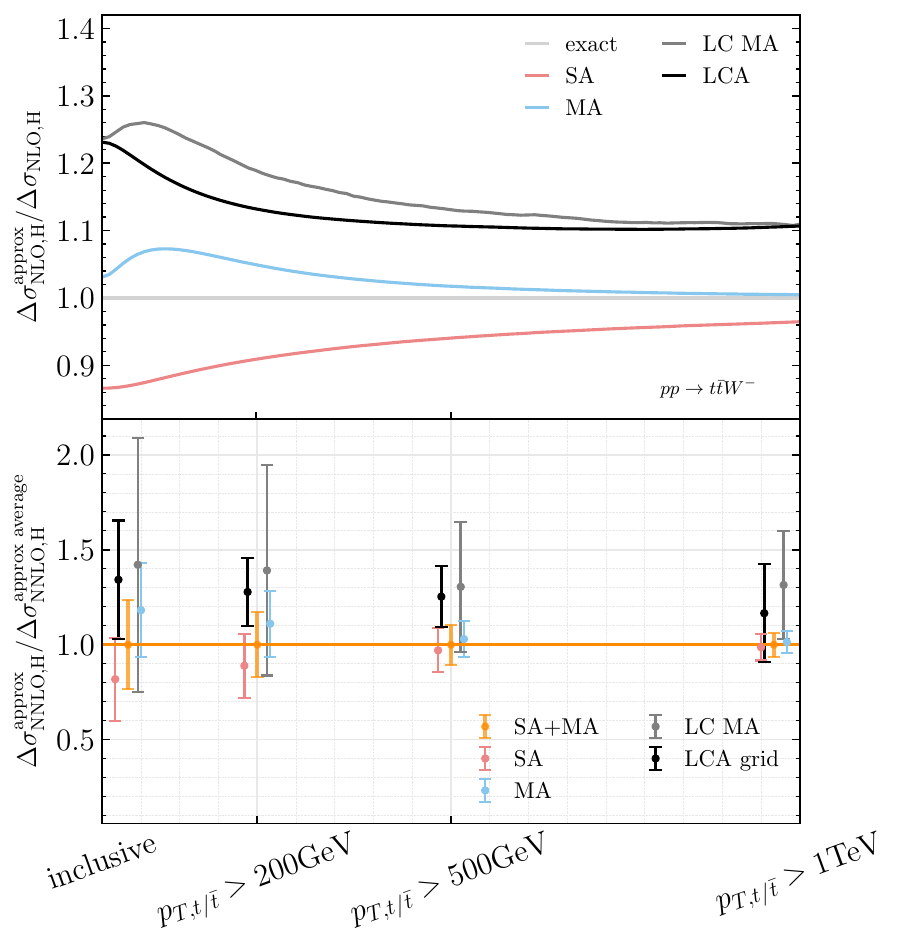}
  \caption{
Results for $\Delta\sigma_{\rm NLO,H}$ (upper panel) and $\Delta\sigma_{\rm NNLO,H}$
(lower panel) in the case of $\ttWm$ production, computed using
the various approximations (SA, MA and LCA), for different
cuts applied to the transverse momenta of the top quarks.
At NLO the approximations are normalised to the exact result, while
at NNLO to the best result (SA+MA) from Ref.~\cite{Buonocore:2023ljm}.
The uncertainties assigned to each approximation at NNLO are discussed in the text.
Similar results are obtained for $\ttWp$.}
  \label{fig:comp}
\end{figure}

In Fig.~\ref{fig:comp} our new NLO and NNLO results for $\Delta\sigma_{\rm (N)NLO,H}$ based on the LC approximation (LCA) are compared to those of Ref.~\cite{Buonocore:2023ljm}, as functions of the lower cut applied to the transverse momenta of the top quarks, $p_{T,t/{\bar t}}$.
The {\it soft approximation} (SA) is obtained by evaluating the $\ttW$ amplitude in the soft $W$-boson limit.
The {\it high-energy} approximation, instead, is formally valid in the limit where the top-quark mass is much smaller than all other relevant energy scales. It is constructed via the {\it massification} procedure~\cite{Mitov:2006xs,Wang:2023qbf} and thus labelled MA in the following.
We stress once more that the approximations used throughout this Letter are applied only to the hard-virtual coefficient $H^{(n)}$ \mbox{($n=1,2$)} in Eq.~(\ref{eq:Hn}); the Born cross section ${\rm d}\sigma_{\rm LO}$ appearing in Eq.~(\ref{eq:sigmaH}) is always computed exactly.

We start by considering the upper panel of Fig.~\ref{fig:comp}, where the various NLO approximations are compared to the exact one-loop result.
As discussed in Ref.~\cite{Buonocore:2023ljm}, the SA (MA) underestimates (overestimates) the exact prediction, and both approach the latter when a hard cut is imposed on the top-quark transverse momentum. Indeed, the large-$p_{T,t/{\bar t}}$ region corresponds to a kinematic configuration in which both SA and MA are expected to reproduce the full amplitude. In this regime, the agreement with the exact result is particularly good for the MA, with differences at the percent level.
The LCA, on the other hand, consistently overestimates the exact result, with a deviation of about $22\%$ at the inclusive level, which decreases to roughly $10\%$ in the high-$p_{T,t/{\bar t}}$ region. We further observe that, at large $p_{T,t/{\bar t}}$, the LCA result is in very good agreement with its high-energy approximation (LC MA), mirroring the behaviour observed for the full-colour prediction.

We now turn to the lower panel of Fig.~\ref{fig:comp}, which shows the various approximations to the NNLO hard-virtual contribution, $\Delta\sigma_{\rm NNLO,H}$, normalised to the reference prediction of Ref.~\cite{Buonocore:2023ljm} (in orange). The latter was defined as the arithmetic average of the SA (in red) and MA (in blue) results, with their uncertainties combined linearly.
In Ref.~\cite{Buonocore:2023ljm}, the uncertainties associated with SA and MA were estimated under the assumption that the quality of a given approximation of $\Delta\sigma_{\rm NNLO,H}$ cannot be expected to exceed the accuracy observed at NLO. More precisely, the relative uncertainty on $\Delta\sigma_{\rm NNLO,H}$ was taken to be twice the relative difference between the approximate and exact predictions for $\Delta\sigma_{\rm NLO,H}$. This conservative prescription reflects the fact that the two approximations employed in Ref.~\cite{Buonocore:2023ljm} have a {\it dynamical} origin and are applied well beyond their nominal region of validity, requiring suitable kinematic projections. The final uncertainty further includes variations of the subtraction scale $\muIR$ by a factor of two around the central scale choice $Q$.

In the present calculation, the two-loop amplitude is evaluated within the LCA.
Unlike SA and MA, the LCA is a {\it parametric} approximation and does not rely on any extrapolation or kinematic projection.
Its main limitation is expected in the threshold region, where subleading-colour (SLC) effects are dominant. This region, however, has a negligible impact on the hadronic $\ttW$ cross section. Moreover, a comparison of the exact and LCA results for the one-loop hard-virtual coefficient $H^{(1)}$ reveals no significant distortions (beyond the threshold region) across phase space, and a similarly stable behaviour is observed when comparing the LCA predictions for $H^{(2)}$ and $H^{(1)}$.
Motivated by these observations, we estimate the LCA uncertainty at NNLO from the relative difference between the exact and LCA results at NLO. We also consider variations of the subtraction scale $\muIR$ around the central scale $Q$, but those effects are consistently subleading. The final uncertainty assigned to $\Delta\sigma^{\rm LCA}_{\rm NNLO,H}$ is obtained by linearly combining the LCA uncertainty with an estimate of the numerical error associated with the grid interpolation. The latter is below the percent level inclusively, but increases to about $12\%$ in the high-$p_{T,t/{\bar t}}$ region.
The resulting predictions for $\Delta\sigma^{\rm LCA}_{\rm NNLO,H}$ are shown in the lower panel of Fig.~\ref{fig:comp} as black markers, together with the LC MA results shown in grey. The latter are obtained from the LC two-loop massless $W+4$-parton amplitudes~\cite{Abreu:2021asb,Badger:2021nhg}, as in the MA case, but with the massification ingredients retained in LC.
We find that the LCA prediction is systematically larger than the SA+MA result of Ref.~\cite{Buonocore:2023ljm}, while remaining compatible with it within the estimated uncertainties. 

\renewcommand{\arraystretch}{1.5}
\begin{table}[]
\centering
\begin{tabular}{|c|c|c|}
\hline
        $\Delta\sigma_{\rm NNLO,H}$[fb] & $t{\bar t}W^-$ & $t{\bar t}W^+$\\
\hline
	SA & $15.67\phantom{0}(0)\pm 4.21$ & $34.59\phantom{0}(0)\pm \phantom{0}9.18$\\
\hline
        MA & $22.62(15)\pm 4.78$ & $49.87(40)\pm \phantom{0}8.74$\\
\hline
        SA+MA & $19.14\phantom{0}(8)\pm 4.49$ & $42.23(20)\pm \phantom{0}8.96$\\
\hline
	LCA & $25.70\phantom{0}(0)\pm 5.68$ & $56.76\phantom{0}(0)\pm 12.03$\\
\hline
\end{tabular}
\caption{Contribution of the hard-virtual coefficient $H^{(2)}$ to the inclusive NNLO cross section in the various approximations (SA, MA and LCA).
The numerical error from the Monte Carlo integration is shown in brackets. The additional error denotes our estimate of the uncertainty of each approximation.
The LCA uncertainty also includes an estimate of the interpolation error.}
\label{tab:H2}
\end{table}

Finally, the numerical results for the inclusive NNLO hard-virtual contribution are reported in \cref{tab:H2}.
As shown in \cref{fig:comp}, the LCA results are consistent with the SA+MA prediction of Ref.~\cite{Buonocore:2023ljm}, with slightly larger uncertainties.
Combining the new result for the two-loop contribution with the remaining NNLO contributions~\cite{Buonocore:2023ljm}, which are evaluated without any approximation, we obtain the prediction for the NNLO QCD cross section reported in \cref{tab:xs} (second row). For comparison, we also report the NNLO QCD result of Ref.~\cite{Buonocore:2023ljm}.\footnote{The tiny difference between $\sigma_{\rm NNLO}^{\rm SA+MA}$ and the values given in Table~I of Ref.~\cite{Buonocore:2023ljm} is due to the correction of a bug in the transverse-momentum soft function of Ref.~\cite{Devoto:2025eyc}.}

\renewcommand{\arraystretch}{1.5}
\begin{table}[]
\centering
\begin{tabular}{|c|c|c|}
\hline
        $\sigma_{\rm NNLO}$[fb] & $t{\bar t}W^-$ & $t{\bar t}W^+$ \\
\hline
	SA+MA & $235.4(0)^{+5.1\%}_{-6.6\%}\pm 1.9\%$ & $474.9(2)^{+4.8\%}_{-6.4\%}\pm 1.9\%$ \\
\hline
	LCA & $241.9(0)^{+6.4\%}_{-7.3\%}\pm 2.3\%$ & $489.4(1)^{+6.3\%}_{-7.2\%}\pm 2.5\%$ \\
\hline
\end{tabular}
\caption{NNLO QCD cross section based on the LCA for the two-loop amplitude (second row), compared with the results of Ref.~\cite{Buonocore:2023ljm} (first row). The perturbative uncertainties are estimated through standard seven-point scale variation, while the additional uncertainty assigned to $\sigma_{\rm NNLO}^{\rm LCA}$ takes into account the missing SLC contributions and the grid-interpolation error, as discussed in the main text.}
\label{tab:xs}
\end{table}

The new $\sigma^{\rm LCA}_{\rm NNLO}$ results are about $3\%$ higher than the $\sigma_{\rm NNLO}^{\rm SA+MA}$ predictions from Ref.~\cite{Buonocore:2023ljm} but consistent with them within the stated uncertainties.
Comparing with the results in Table~\ref{tab:H2} we see that the impact of the LC two-loop virtual contribution is rather large: it amounts to about $11\%$ of the NNLO cross section.

\section{Summary}

In this Letter, we have presented an NNLO QCD calculation of the inclusive $\ttW$ cross section at the LHC. For the first time, the two-loop virtual amplitudes have been computed explicitly in the generalised leading-colour approximation. As a \mbox{$2\to 3$} process involving multiple mass scales, $\ttW$ production represents one of the most challenging applications accessible to available multi-loop amplitude techniques, and the calculation reported here pushes these methods to their present limits.

Our new results are in full agreement with the predictions of Ref.~\cite{Buonocore:2023ljm}, which were obtained with two dynamical approximations of the two-loop contribution. 
Since those predictions currently serve as a benchmark for comparisons with experimental measurements, this validation constitutes a particularly valuable and relevant result of this work.
We estimate the uncertainties of the NNLO cross section due to the missing subleading-colour contributions at the $2.5\%$ level --- slightly larger than the estimated uncertainty of the dynamical approximations of Ref.~\cite{Buonocore:2023ljm}.
This estimate follows from the expected ${\cal O}(20\%)$ parametric accuracy of the LC approximation, together with the sizeable impact of the two-loop contribution on the NNLO cross section for this process.

Beyond its phenomenological implications, this work also constitutes an important methodological milestone, marking the first application of our approach --- based on the numerical evaluation of the rational coefficients of the special functions entering the two-loop amplitude --- in a complete phenomenological calculation.
This achievement paves the way for NNLO QCD predictions for other important and complex processes.

Our results also open the door to further improvements.
In the short term, our LC calculation may be combined with an approximate treatment of the subleading-colour contribution, following the strategy of Ref.~\cite{Buonocore:2023ljm}, for which the basic ingredients are now available.
In the longer term, the inclusion of exact full-colour contributions will require further methodological advances.

\vspace{0.5cm}
\noindent {\bf Acknowledgments}\\
\noindent
M.G., S.K.\ and C.S.\ would like to thank Luca Buonocore, Simone Devoto and Luca Rottoli for comments on the manuscript.
This work has been supported by the European Research Council under the European Union’s \textit{Horizon 2020} and \textit{Horizon Europe} research and innovation programmes through grants No.~101040760 (ERC Starting Grant \emph{FFHiggsTop}: M.B., D.C., V.C., T.P., and M.P.), 
No.~101167287 (ERC Synergy Grant \emph{MaScAmp}: V.C.), No.~949279 (ERC Starting Grant \emph{HighPHun}: M.D., S.D., and L.T.), No.~101044599 (ERC Consolidator Grant \emph{JANUS}: M.D.), and No.~101118787 (ERC Synergy Grant \emph{UNIVERSE PLUS}: S.D.).
Views and opinions expressed are however those of the author(s) only and do not necessarily reflect those of the European Union or the European Research Council Executive Agency. Neither the European Union nor the granting authority can be held responsible for them.
The work of X.C., M.G., and S.K.~is supported by the Swiss National Science Foundation (SNSF) under contract 200020$\_$219367.
X.C.~is partly supported by the UZH Postdoc Grant, grant No.~[FK-25-104].
The work of C.S.~is partly supported by the Excellence Cluster ORIGINS, funded by the Deutsche Forschungsgemeinschaft (DFG, German Research Foundation) under Germany’s Excellence Strategy~---~EXC-2094-390783311.
S.Z.~was supported by the Swiss National Science Foundation (SNSF) under the Ambizione grant No.~215960.

\bibliography{bibliography}

\end{document}